**ABOUT DIFFUSION IN POROUS MEDIUM: the role of the correlation length**

P.C.T. D´Ajello, G.L. Nunes, J.J. Piacentini and L. Lauck

Universidade Federal de Santa Catarina

Departamento de Física/ CFM

Brazil

P.O. Box 476 – CEP 88040-900

Fax 55 (48) 3721 9946

e-mail: pcesar@fisica.ufsc.br

**ABSTRACT**

In this paper we develop a model to describe the diffusion process in a porous medium. For the observed decrease in current yield, we propose other causes than difference in diffusivity, which we consider unaltered by the porous medium. The physical situation we try to model consists of systems of reduced dimensions ($\sim 0.001 - 1.0\ cm^3$) with pores of sub micrometric dimension. This is particularly suitable to represent organic structures or special cells in electrochemical devices. We try to explore two basic contributions as an answer for diffusion fading in porous medium, namely, the effect of the void geometry and a dissipative process as well. This dissipative process is in the kernel of our analysis and it is related to the heterogeneous fluctuations of the flux lines occurring at the border among pores. To mimic biophysical and electrochemical conditions we also include in our model a reactive process, such migration of species do not need, necessarily, a pressure gradient because of the reaction diffusion process. As usual, the mathematical description of the phenomenon is introduced through the prescription of boundary and initial conditions for a transport equation, however here we decouple the geometry from the reaction-diffusion-convection problem examining a unitary conduct, in which the fluid flows and that, replicated in the space, is able to generate the entire system. We estimate the current transient provided by our model and compare it to the one obtained by the process of controlled growth of metallic inverse opals prepared by electrochemical crystallization of matter, so the theoretical results can be compared with experimental data.

## 1. INTRODUCTION

Diffusion and reaction, and flow of matter in porous medium, compose a very appellative subject because it plays a central role in questions like the extraction of oil from deep wells, the drain of water by soil and the filtering of water toward the aquifer, as well as the delivering of glucose and oxygen from the vascular system to brain cells [1-5]. Unfortunately, the prescription given by hydrodynamics to solve the most general form of this question requires, for an ideal fluid, the solution of a set of coupled equations, namely the Euler's equations, the equation of continuity and also an equation of state. Since, under these conditions, analytic solutions are unreachable, a general approach has being used to circumvent mathematical entanglements. It consists in assuming that the only effect of the porous medium on the diffusion process is to alter the diffusion constant. Nicholson [4,5], by an ingenuous mathematical argument estimated this new diffusion coefficient to be,

$$D^* = \frac{\varepsilon \delta}{\tau} D,$$
(1)

where the diffusion constant in free space $D$ is multiplied by a factor (lower than one) that includes the porosity $\varepsilon$, the constrictiveness $\delta$ and the tortuosity $\tau$. These three constants are related to the geometry of the porous medium, as one can see in the work of Nicholson. We argue that such an argument has two severe flaws: first, the used average procedure ruled out any fluctuation, which could be associated with flow in porous medium. Second, it regards the geometry as having an influence on the diffusivity. This type of approach introduces the interpretation that all the reduction in the observed mass transport in porous media is solely a consequence of the geometry of the medium. In fact, the theory says that the flow of matter is still described by the same equations but the diffusivity in the porous system is different from that found in a free medium. In any way the inclusion of the geometric contribution is essential and any model wishing to describe diffusion, convection and reaction in a porous system should take it into account. However, in systems characterized by very small dimensions, such as those observed in electrochemical plating process, some sort of fluctuations could be relevant to describe the early evolution in current transients.

In this article we develop a model to describe the current transients observed by Sapoletova et al. [6], which developed an experiment to examine the kinetics of nickel electrodeposition through a template of ordered polystyrene spheres, a system that can be considered a prototype of a porous system of reduced dimensions ($m$ layers of monodisperse polystyrene microspheres with an average diameter of 530 nm in a sample that shows electroactive area of $0.25 - 1\ cm^2$). In figure 4 of their article Sapoletova et al. [6] shows a current transient registered in the course of Ni electrodeposition through the porous lattice. The current profile shows an initial peak followed by others, with amplitudes that decrease following the oscillations of the holes defined by the stacking of the spheres.

To model this system we search for an approach quite different from that followed by Nicholson [4,5]. We try to decouple geometry from the reaction-diffusion-convection problem. Thus we assert the way the geometric contribution arrives into the description has no dependence with the coefficients that characterize the flow like diffusivity, reaction constant, etc. On the other hand this independence is convenient because it allows for the introduction of fluctuations on the path followed by any volume element of the fluid through the meandering roads followed in its flow. This is our first goal in this article, to reveal fluctuations in the flux´s lines. They can be schematically represented by the sketch in Fig.1 where we show a particular pathway (a duct) through which the fluid and/or the particles flow. This duct gains or losses particles because the random fluctuation of the path lines on regions that connect the duct to next near neighbors ones. The second goal of this article is to introduce the geometric shape of the voids, the pores, through a boundary condition, related but not mixed with the random fluctuations of the stream lines or diffusivity. With these ideas we try to put forward an alternative description of transport of species guided by the concurrent contribution of diffusion, convection and reaction in porous medium. We hope this approach turns out to be useful as a counterpoint to the standard description, well represented by the aforementioned works of Nicholson et al. [4,5].

Our model could be applied to a large number of situations as indicated at the beginning of this introduction; however, to give some materiality to our speculation, we choose to describe a particular experimental circumstance. In the experimental array whereupon we focus our attention, the porous system is composed of submicrometer spheres in an fcc structure, on the top of a flat electroactive substrate. An electrolytic medium fills the interstices among the spheres and also contains the species which we observe to migrate through the pores toward the electroactive substrate, where they react producing a material deposit that grows from its top and fills the interstices among the spheres.

In what follows we introduce the model, we discuss the results provided by our model and compare it to the ones observed in the aforementioned experiment we also improve our model analyzing the physical meaning of the diffusion length concept in open systems, and the implications of an increasing correlation length connected with the mean square deviation of the parameter that represents the random fluctuation of the flux' lines. These ideas lead us to use a cluster approximation to describe the current arriving from the contribution of all the units (the ducts) that compose the system.

## 2. GROWTH OF METALLIC INVERSE OPALS BY ELECTRODEPOSITION

In what follows we try to show how a realistic system, formed by a close-packed array of monodisperse colloidal spheres could be reduced to a problem with cylindrical symmetry, amenable to analytical calculations.

In fact the theoretical model focus attention on species, diluted in a liquid medium that fills the space among the spheres ordered in a sequence of monolayers from a flat reactive substrate. The colloidal crystal formed in this way is identified as a prototypical porous medium and has its (111) axis oriented perpendicularly to the substrate, as shown in the top view of Fig. 2(a). Along the (111) direction there is no straight path to the bottom and the ions are forced to deflect laterally in order to reach down the next layer of spheres. In Fig. 2(a) the curved white arrows single out one possible diffusion path for an ion that strikes the location marked with a star. The same diffusion path is shown by black arrows in the schematic vertical cross section of the porous structure (Fig. 2(b)). This particular path will be modeled as a staircase cylindrical vessel. At every inflection point along the twisted cylinder the diffusing particle may either remain in the same vessel or diffuse out into one of three other neighbouring units, which is indicated by the dotted white arrows in Fig. 2(b). Equally, at each inflection point a cylinder may receive particles coming from three other vessels. The porous structure, as a whole, can be seen as a periodic replication of a twisted cylinder, conveniently placed along the $x$ and $y$ directions, as pictured in Fig. 2(c). As we see, the essence of the lateral diffusion of ions is to allow for exchange of particles among neighbouring vessels and this will be described by a periodic function $g(z)$ that modulates, along the length of the cylinder, the intensity of influx/outflux of particles across the walls, plus a random variable that defines the flux direction at each point of exchange. With these considerations two essential features of the problem are taken into account, namely, the lateral diffusion of ions and its periodic and random nature. Therefore, the twisted shape of the cylinders may now be neglected and the problem reduces to an array of porous and straight cylinders directed along the $z$ axis, with a lateral periodicity that mirrors the hexagonal lattice seen at the surface of the porous structure.

The next point of analysis concerns the description of the internal wall of the cylindrical vessel. In real systems the cross-sectional area of a diffusion path has a complex shape. Along the $z$ direction its open portions, where exchange of particles is possible, are periodically intercalated with closed portions of minimal cross-sectional area that are delimited by the closest approximation among three spheres. Moreover, the lateral exchange of particles has an angular dependence with a trigonal symmetry. The model, however, will ignore those details and focus solely on the periodicity of the cross sectional area exhibited along the $z$ direction, assuming for it a cylindrical symmetry, with no angular dependence in the horizontal plane. The internal cross sectional radius of the vessel will vary, in a continuous and periodic fashion, from a minimum value $R_{min}$ up to $R$, which is the external radius of the cylinder, as shown in Fig. 2(d). The need of assuming a corrugated internal surface for the cylinders will become more evident later on.

After this picture we are ready to formulate the mathematical description of diffusion, convection, reaction problem for one single cylindrical vessel. This cylindrical vessel now could be pictured in a standard way with boundaries defined by rigid walls but showing a graded permeability or as a family of profiles defined by the flux lines that conforms migration paths for the particles, the outer of which could fluctuate, also with a gradual and oscillatory intensity like the permeability in the first case.

We look for an expression for the current produced by the reduction and deposition of ions at the electrode surface. The electrode surface lies at the bottom of a cylindrical cavity. The cylindrical cavity is filled by an electrolytic solution that contains the ionic species, homogeneously diluted at the initial time. The ionic concentration inside the finite cylindrical region obeys the balance equation:

$$\frac{\partial c}{\partial t} = D \left[ \frac{\partial^2 C}{\partial r^2} + \frac{1}{r} \frac{\partial C}{\partial r} + \frac{\partial^2 C}{\partial z^2} \right] - \vec{v} \cdot \vec{\nabla} C, \tag{3}$$

where, $C = C(r, z, t)$ is the species concentration, which do not depend on the angular variable due to our assumption of rotational symmetry. In Eq. (3), $D$ is the diffusion constant and $\vec{v}$ the convective velocity, defined in the sense given by the Darcy's law, that is,

$$\vec{v} = v_c \hat{k}. \tag{4}$$

The model is completed by the following initial and boundary conditions:

$$C(r, z, 0) = c_b, \qquad (0 \le r \le R, z \ge 0) \tag{5a}$$

$$C(r, 0, t) = (c_b - c_s)e^{-kt} + c_s, \qquad (0 \le r \le R, \forall\, t) \tag{5b}$$

$$C(r, h, t) = c_b, \qquad\qquad (0 \leq r \leq R, \forall\, t) \qquad\qquad (5c)$$

$$\left.\frac{\partial c}{\partial r}\right|_{r=0} = 0, \qquad\qquad (\forall\, z, \forall\, t) \qquad\qquad (5d)$$

$$\left.\frac{\partial c}{\partial r}\right|_{r=R} = -\alpha c_b (1 - e^{-\nu t}) g(z)\,. \qquad (\forall\, z, \forall\, t) \qquad\qquad (5e)$$

In Eqs.(4-5), $v_c$ is the magnitude of the convective velocity, $c_b$ the constant and homogeneous concentration in the cylinder cavity at the initial time, before the applied electric potential, which allows for reduction of ions on the electrode, is turned on. $c_s$ is the limit concentration of species on the electrode; $k$ is the reaction rate, it describes the rate at which the surface transfers electrons to reduce the ions. In fact it is a function of the electric potential difference between the electrode and liquid medium [7-11] which is taking as a constant due the assumption of a fixed electric potential difference. Therefore $k$ is a factor that defines the reaction kinetics. Despite its simplicity, boundary condition (5b) plays a central role in our description because it defines the concentration drop at the electrode where the ions are reduced and withdrawn from the liquid medium. In cases where $v_c = 0$, if $k = 0$ there is no reaction and there are no concentration changes at the surface ($C = c_b$ at any time). For $k \neq 0$, the concentration at the electrode surface evolves to a stationary value (when $t \rightarrow \infty$). A concentration gradient is produced and ions migrate from the bulk of the solution toward the surface. Boundary condition (5c) guarantees that we are working with a cylinder whose length is at least equal to the stationary diffusion layer h, which defines a distance from the electrode surface, beyond which the ion concentration is assumed to be constant and equal to the bulk concentration $c_b$. Boundary condition (5d) is a consequence of the rotational symmetry of our system. Finally, the time dependent boundary condition (5e) determines the evolution of the concentration of ions that flow through the lateral surface of the cavity ($R$ is the outer cylinder radius), in its normal direction. This boundary condition also depends on $z$, that is, the flux of matter through the cavity's lateral surface can change along its length. This boundary condition contains the parameter $\alpha$ (lower than one), which quantifies the magnitude of the matter flux that crosses the cylindrical lateral surface. Its signal determines the direction of matter flow, inward for negative $\alpha$ or outward if $\alpha$ is positive. The time dependent expression contained inside the brackets is a mathematical requirement to guarantee consistency of the boundary and initial conditions. Thus, at $t = 0$, there is no flow and the system is characterized by a constant and homogeneous distribution of matter $(C(r, z, t) = c_b)$. When the potential is switched on, the symmetry is broken off. Ions begin to react at the cylinder bottom and a gradient arises so that species flow towards the electrode surface. In addition, the flow across the lateral area of the cylinder obeys a transient rule, quantified by the magnitude of the rate constant $\nu$ that appears on the exponential argument of Eq. (5e). A physical reasoning to justify the temporal dependent term in boundary condition (5e) is that $\nu$ quantifies the time interval elapsed until the flow assumes a magnitude that allows for the fluctuation of matter on the border of the cylinders according to the rule prescribed by $g(z)$ and $\alpha$ on the lateral surface of the cavity. Regardless of this interpretation, we wish to emphasize the relevance of boundary condition (5e) for our model. It is the condition that sustains the flexibility of the model, that is, its capability to reproduce different situations, according to the sign of $\alpha$ and the form of the function $g(z)$. Thus, good choices of $\alpha$ and $g(z)$ make the model more real.

To solve Eq. (3) we need to implement two transformations which are shown in the Appendix. The solution for the charge current that flow through the reactive area is given by (see appendix):

$$\begin{aligned}
I(0, t) = -D\bar{z}F\pi R^2 \Bigg\{ &(c_b - c_s)e^{-kt}\left(\beta - \frac{1}{h}\right) + c_s\left(\beta + \frac{1}{h}\right) + \frac{c_b}{h}e^{-\beta h} \\
&+ \frac{2}{h}c_b\beta^2 \sum_{n=1}^{\infty} \frac{\omega_n}{\beta^2 + \omega_n^2}\left(\cos(n\pi)e^{-\beta h} - 1\right)e^{-\mu_n t} + \frac{2}{h}c_b\sum_{n=1}^{\infty}\Omega\frac{\cos(n\pi)}{\mu_n}e^{-\beta h}(1 - e^{-\mu_n t}) \\
&+ \frac{2}{h}(c_b - c_s)\sum_{n=1}^{\infty}\frac{(\Omega - k)}{\mu_n - k}\left(e^{-\mu_n t} - e^{-kt}\right) + c_s\Omega\sum_{n=1}^{\infty}(1 - \cos(n\pi))\frac{(1 - e^{-\mu_n t})}{\mu_n} \\
&+ \frac{2}{h}\frac{D\alpha c_b}{R}\sum_{n=1}^{\infty}2g_n\omega_n\left[\frac{(e^{-\nu t} - e^{-\mu_n t})}{\mu_n - \nu} - \frac{(1 - e^{-\mu_n t})}{\mu_n}\right]\Bigg\}.
\end{aligned} \tag{6}$$

In Eq. (6):

$g_n = \int_0^h g(z)\sin(\omega_n z)\,dz$,

$\beta = \frac{v_c}{2D}$,

$$\Omega = \frac{v_c^2}{4D},$$

and

$$\mu_n = \frac{Dn^2\pi^2}{h^2} + \frac{v_c^2}{4D}.$$

Because the convective velocity $v_c$ is not the velocity of a volume element in the fluid but the flow velocity through the porous medium, we could invoke the Darcy's law to make explicit its dependence with some pressure gradient, that is;

$$v_c = -A\frac{\partial}{\partial z}P, \tag{7}$$

where, $A$ is a constant (usually related to the porosity, the permeability and the viscosity) and $\frac{\partial}{\partial z}P$ is the pressure gradient.

When there is no convection, $v_c = 0$, and the concentration of species on the electrode surface goes to zero, $c_s = 0$, we are in fact working with heterogeneous diffusion of species in a liquid quiescent medium and Eq. (6) reduces to the case we analyzed in reference [11].

To describe the charge current for the system defined in Fig.2 we must take into account the variation of the cross sectional area of the cavity, which is a periodic function of $z$ and that changes with time due to the systematic matter deposition. This cross sectional area is a periodic function of $z$ and described by $\pi\tilde{R}^2$, as sketched in Fig. 2d.

$$\pi\tilde{R}^2 = \pi\left(R - 0.25R\left[1 - cos^2\left(\frac{\pi}{2R}z\right)\right]\right)^2. \tag{8}$$

To introduce the geometry in our description we define the function $g(z)$, which we assume as:

$$g(z) = \gamma cos^2\left(\frac{\pi}{2R}z\right) \qquad 0 \le z \le mR. \tag{9a}$$

$$g(z) = 0 \qquad z > 2mR. \tag{9b}$$

Eqs. (9) represents an array composed of $m$ monolayers, produced by close packed spheres, above which the fluid fills the free space volume. In Eq. (9a) we introduce $\gamma$, a constant , to get $g_n$ assume values closer to the unit.

To describe the complex dynamic behavior of a real porous system, where influx or outflux of particles occurs at random, on the lateral surface of our corrugated cylinder, we should consider a statistics for $\alpha$, for which we adopt the simplest proposition. We assume that the sign and the magnitude of $\alpha$ are random variables with a null ensemble average, assigning to $\alpha$ a plain random walk behavior. Thus, the change in flow's intensity and direction through the lateral surface of the cavity is represented by the product of functions $\alpha g(z)$ that defines the representative horizontal bars appearing on Fig. 2d. The random variation in $\alpha$ is achieved by a Monte Carlo algorithm introduced in Eq. (6). We, in fact, are considering just one single corrugated vessel, thus fluctuations on $\alpha$ implies the influx and outflux of particles to this vessel which corresponds to a particular path for a particle. To take into account this lateral flux the following prescription is taken: the initial value is $\alpha_0 = 0$. Then, at a certain instant $t_1$ a constant increment $\pm\Delta\alpha$ is chosen with the sign taken at random, which defines a new value for $\alpha$, $\alpha_1 = \alpha_0 \pm \Delta\alpha$, that is maintained for a time interval $\Delta\tau$. The procedure is repeated until the current transients are concluded.

There is a final aspect to consider before the application of the method. Because the heterogeneous reaction that occurs at the electrode results in a material deposit at the bottom, there will be a progressive increase in the deposited film thickness and consequently a progressive shift on the g(z) function that mimics the lateral flux. This effect changes the flux geometry near the topmost deposited layer. To account for that, the reactive surface is kept fixed at $z = 0$ whereas $z$ is shifted by a discrete and constant quantity, so that we have $g(z')$ with $z' = z + \Delta z$ where $\Delta z$ is the increment on the deposit thickness after a given time interval $\Delta t$. This trick allows us to compute Eq. (6) at $z = 0$ taking into account the correction in $g_n(z)$.

Fig. 3 presents the current transients obtained for a system composed by four monolayers of spheres. Eq. (6) is used to generate the current profiles, using the standard values $D = 1x10^{-6} cm^2 s^{-1}$, $c_s = 0$ and $k = 0.89 s^{-1}$. We also assume the cylinder radius is $R = 300 nm$ and the magnitude of the depletion layer $h$, i.e., the perpendicular distance from the reactive surface, beyond which the concentration of species remains constant is $h = 3x10^{-3} cm$. Each curve in Fig. 3 corresponds to a different value for the convective velocity, so that we could appreciate the effect of increasing the convective velocity (pressure gradient). In Fig.3 the curves with positive convective velocity have no appreciable physical significance. They were presented only to indicate the expected behavior if it were possible to remove the particles from the vicinity of the electrode, moving them with the velocity $v_c$.

Before continuing we should note that the whole system is formed by many vessels of this kind and the current transients, as obtained by Sapoletova [6], Sumida [12] and Szamocki [13], who worked with similar systems, correspond to the overall current that includes all vessels that composes the system. Besides, on the electrode, where just a collection of dispersed points are electro active at the initial time, constituting themselves the nucleation points on which reaction occurs, but belonging to vessels located at a distance from each other, we expect that on the beginning the total current be equal to the current we have just calculated for one single vessel multiplied by the number of nucleation points on the electrode surface. See Fig. 4a for an image.

Up to this stage the total current is been taken as a sum of contributions that came from uncorrelated units, but when time evolves, and a nucleation point develops its hemispherical form and propagates its activity to the bases of the near neighbor ones, then the vessels turn to be correlated through the interchange of matter on the borders among them. Under this circumstance we must search for a more adequate proposition in order to compute the total current.

In the next section we will examine the effects of correlation among vessels and how that it competes to produce the attenuation of matter transport in porous systems.

## 3. THE FLUCTUATIONS AND THE GEOMETRIC EFFECT.

Up to now we considered the current transient occurring in just one corrugated vessel. The current profiles are then described by Eq. (6) with geometrical contribution included through the proposition of $g(z)$, which defines a shape for the vessels, and also through $h$ that enters to define the vertical boundary in our finite system. Fluctuation effects are included into the solution through the definition of the random variable $\alpha$ that quantifies the magnitude and direction of the lateral flux through the vessel border. However, two basic questions arise when we are obliged to consider all the vessels that make up the system. One still concerns the description of the flux in just one vessel and is related to the changes produced in the current yield due to variations in on the magnitude of $h$ as a consequence of the thickening of the deposited layer. The second is an effect that appears only when we consider the whole system with the inclusion of all activated vessels. This effect is related to the correlation of the flux among vessels, following the fluctuations at its borders and the propagation of this effect from there. Let us first consider the meaning of $h$ and its implications when considering the porous system.

By its very definition, $h$ represents the thickness of the diffusion layer. It enters into our description through the boundary condition (5c) working as a threshold magnitude to assigns the region which is affected by diffusion (where there is a concentration gradient). Because of that, it is natural to expect a change on the magnitude of $h$ if two or more layers of monodisperse spheres are added to the system and it is natural as well to expect a change on the current yield if there is an increase in the number of monolayers that constitute the system. In fact, in a reaction-diffusion process $h$ results from the competition between these kinetics. Thus, if diffusion kinetics (which feeds the species to be reduced on the reactive surface) is higher than the reaction kinetics (which accounts for the withdrawal of species on the reactive surface), then $h$ is lower than it would be if the reaction kinetic was higher (due a strong reaction rate $k$) than the diffusion kinetics (which depends on the diffusion constant $D$). Of course, $h$ depends on the concentration of the reacting species, but if concentration is set fix, then the interplay between diffusion and reaction defines $h$ for an initially homogeneous, free and non convective medium.

To make easier our understanding, let's assume a diffusion kinetics much higher than the reaction kinetics, so if a particle is withdrawn from the medium by reaction, the system recovers homogeneity (diffusion is faster than reaction) well before a second reaction may occur and in this case $h = 0$. Now, if the reaction kinetics is faster than diffusion the homogeneity near the reacting surface is lost within a layer which grows until a steady state is reached, whereupon defining a value to $h$.

However, the relevant question here is about the changes produced on $h$ once the concentration, the diffusivity and the reaction rate are fixed, but there is a continuous change in the relation volume-reactive area. Let's consider first the simplest case, that of a cylindrical cavity with a reactive base and filled with an electrolytic solution. This system has a well defined $h$ according the relation between diffusivity and reactivity in free space. Now we may ask: does the thickness of the diffusion layer $h$ change if we introduce some balls within the cylindrical cavity? Or, which is the same; does the thickness of a porous medium affect $h's$ magnitude? The answer is yes. Because the number of particles available to feed the reaction is proportional to the density, which does not change by the inclusion of the bead, but the volume near the reactive substrate changes producing a reduction in the number of particles available to be reduced on it. Then volume plays the role of the diffusivity if the reacting area is set constant. On the other hand, the reactivity is proportional to the effective area through which the particles are suppressed from the solution, so area plays the role of the reaction rate. As a consequence, $h$ should increase after a reduction of the effective volume, as well as increase with the reactive area. Then, if $h_o$ is the diffusion layer thickness for a right circular cylindrical and free cavity, defined by a volume $V_o$ and a reactive area $A_o$ then, the effective magnitude of the diffusion layer thickness for the same system, but with some balls inside the cylindrical cavity, is:

$$h_{eff} = p_v \frac{V_o}{V_{eff}} h_o + p_a \frac{A_{eff}}{A_o} h_o, \qquad (10)$$

where $V_{eff} \leq V_o$ and $A_{eff} \leq A_o$. In Eq.(10) $p_v$ and $p_a$ are weight factors $(p_v + p_a = 1)$ to account for the relative importance of the volume and area rates, respectively. The proposed form for Eq. (10) follows from the independence between the ratios of volume and area, and the connection of the volume rate to the diffusion constant $D$, which is also independent from the reaction rate $k$, which is related to the rate of the reactive areas.

For a cylindrical cavity with radius $R$, height $H$ and no spheres inside $V_o = \pi R^2 H$. If n spheres are introduced into the vessel the volume is reduced, $V_{eff}(t_0) = \pi R^2 (H - n\frac{4}{3}R)$ but the area is almost the same $A_{eff}(t_0) \cong \pi R^2$. The decrease in the volume is proportional to the increase on $h$ and the system with beads shows $h > h_0$. Both, the effective volume and the effective area are now functions of time because a deposit starts to grow from $t_0 = 0$. As an example, let´s consider a particular time $t = t_{1/2}$ that indicates the deposit has grown up to the hemisphere of the lower sphere; at this time $V_{eff}(t_{1/2}) = \pi R^2 \left[ H - \left( 4n - \frac{1}{2} \right) \frac{1}{3} R \right]$ and $A_{eff}(t_{1/2}) \cong 0$. Thus, when time goes on the deposit thickness increases gradually burying the balls, and increasing the effective volume to the limit $V_{eff} = V_o$ and $A_{eff} = A_o$, when all the spheres are buried by the deposit, and with $h$ returning to its initial value $h = h_0$, because the rest of the cylinder is free of beads.

For the corrugated cylinder we consider in this work:

$$A_{eff} = \pi \tilde{R}^2, \tag{11a}$$

with $\pi \tilde{R}^2$ given by eq. (8), and

$$V_{eff} = V_o - \pi R^2 (2nR - h_d) + \frac{\pi R^3}{16} \left[ \frac{81}{8} \left( 2n - \frac{h_d}{R} \right) - \frac{1}{\pi} sin\left( \frac{\pi h_d}{R} \right) - \frac{1}{16\pi} sin\left( \frac{2\pi h_d}{R} \right) \right], \tag{11b}$$

with $h_d$, the height of the deposition layer, ranging from zero to $2nR$. Once the deposit height evolves continually, we must proceed carefully when applying Eq. (6).

Because in our model $h = h_{eff}$ and according the Eq. (6), $I(z,t) \sim h^{-1}$, the striking consequence is the fact that $I(z,t) \to 0$ if $h \to \infty$. This means that the inclusion of a monolayer, in order to make thicker the porous system, will produce a corresponding reduction in the current peak in obedience to the correction introduced by Eqs. (11). If this reasoning is correct, we can define a length $l_i$, which we call infiltration length, that gives the minimum thickness for a porous layer to reduce the flux by two orders of magnitude in comparison to a free system with identical dimensions. In the present case, assuming $p_v = 1$ and $p_a = 0$ we get $\frac{V_0}{V_{eff}} \geq 10^2$, which requires $2nR = \left(\frac{0.99}{0.367}\right) h_o$ when $h_d = 0$. Using the values given to $R$ we get $n = 1348$, that is, we need 1348 monolayers of monodisperse spheres to reduce the current through porous system to $\frac{1}{100}$ of the current for a system with same dimensions but free of beads.

In Fig. 4 we show the current transient generated by the use of Eq. (6) in a four layer system when there is no convection. Curve (a) in Fig. 4 is the continuous line given in Fig. 3, obtained when we do not apply any correction to $h$. Curve (b) is an equivalent profile, defined by the same parameters, also obtained through Eq. (6), but with $h$ corrected according to Eqs. (10) and (11). The comparison does not lead to changes, but we show an inset, where two current profiles are depicted, one for a porous system formed by four monolayers (the black curve) and another formed by just one monolayer(the gray curve). In spite of this small difference in thickness is already possible to notice a reduction in the magnitude of the current intensity for the thicker system as predicted by our reasoning.

Now we will consider the last additional effect, connected to the existence of a time dependent correlation length among corrugated vessels, arising as consequence of the localized fluctuations in the matter flux at the borders of each vessel.

Let us begin with a statement. The current recorded on experiments gives the total current obtained from the superimposed contributions that came from every one of the corrugated vessels filling the space. In this sense, to compute the total current in the simplest way, it is enough to take the current obtained from one single vessel and multiply the result by the number of vessels needed to fill the entire sample. However, this is a poor approximation, because it assumes that the vessels are uncorrelated among them, which is not true, once the role played by the parameter $\alpha$ introduces a fluctuation that competes to mix the flux among neighboring vessels. This assumption also implies that all vessels have an activated base at the initial time. A more appropriated assumption is to consider that the vessels are uncorrelated only at the initial time, when the disturbance introduced by fluctuation, from the edges of pores, has not spread out yet. We simultaneously hope that not all cylindrical cavities have, on their bases, points that are activated at the initial time. Besides, we assume the activation evolves with time starting off from the previously activated ones. This scheme seems plausible if the electrode is a doped silicon layer [14]. If so, when time progresses, the fluctuations originated at the borders of the vessels show a dispersion that correlates the flow among nearest neighboring vessels. In other words, the growth of a correlation length requires that an increasing number of vessels start to show a current that obeys a collective behavior, which could be represented by a cluster approximation.

In Fig. 5 we try to represent the growth of activation points from their distribution at the initial time. Neighbor vessels correlate among themselves with time to produce a current given by the sum of current that flow through each one. However, due to this association, the fluctuations around the mean current is reduced obeying the statistics followed by  . We can appreciate this effect looking at Eq.6, which gives the current in each vessel. Every vessel follows the same prescription, but the last term floats according to $\alpha$ producing a contribution intrinsically different for each vessel, given that each of them follows a particular sequence of random choices of $\alpha$. So, as time goes by, the sequence of random number that generates the magnitude of $\alpha$ differs from one vessel to the other, because generated by different seeds for the random sequences. Therefore, because $\alpha$ is a random zero mean variable, an increase on the number of vessels, used to compute the current, implies a decrease on the current fluctuation, and a lowering on the contribution to the total current coming from the last term in Eq. 6. In the inset appearing in Fig. 6 we show a comparison between the current generated by just one vessel and the mean current given by four correlated vessels.

In order to compute the total current after this perspective we assume the simplest hypothesis. At any time the total current is given by the weighted average coming from the contributions given by the current flowing through the single vessels plus that flowing through the clusters of nearest-neighbor correlated vessels, and that flowing through clusters of next-nearest neighbor correlated vessels and so on. Here we call cluster of first order correlated vessels the ensemble formed by a vessel and all its nearest neighbors (four in the present case). The same reasoning gives us a second order cluster as the one formed by ten completely correlated vessels, because we also include the second neighbors to the cluster. Of course, a detailed description of the currents depends also on a statistics for the growth of clusters but here we use the simplest proposition assuming that at any time there is just two types of clusters, that of order $n$ and those with order $(n + 1)$. In this approach, the total current can be expressed as a linear interpolation which takes into account the contributions from each set of clusters multiplied by the relative weight assigned to them at any time. In fact, we will see that in this case we don't need a better kind of proposition due to the strong attenuation of the fluctuations with increasing orders of the clusters.

To continue we must define: (i) the correlation length, as a time dependent function; (ii) how to compute the contribution of a cluster of correlated vessels and, (iii) the weighted factor used to achieve the total current.

(i) As defined from the beginning, the fluctuations are related to random oscillations of the flux´s lines at the border of the pores (see Fig. 1). A fluctuation in a flux´s line is also a fluctuation on the species that enter or leave a pore in a particular vessel. Besides, the statistical behavior postulated to $\alpha$ was that of a one dimensional Brownian motion that shows dispersion given by,

$$(\Delta x^2) = 4Npql^2 \,,$$

(12)

if we consider the displacement of a Brownian particle after $N$ steps of equal length $l$ with equal probability to move to the right or to the left, $p = q = \frac{1}{2}$. From there we obtain a relation for the diffusion coefficient,

$$D = \frac{4Npql^2}{2t} = \frac{L^2}{t} = \frac{(\Delta x^2)}{2t}.$$

(13)

Where $L$ is a diffusion length.

The correspondence to our case is given through the relations:

$N = \frac{t}{\Delta\tau}$, $\quad l \leftrightarrow \alpha$ and $\overline{(\Delta x^2)} \leftrightarrow \overline{(\Delta\alpha^2)}$. Here $\Delta\tau$ is the time interval between two random choices for the signal of $\alpha$, as explained before. By similarity $\frac{l^2}{t} = 4D \sim \overline{(\Delta\alpha^2)}$, and to recover the correct dimensions,

$$\frac{l^2}{t} = b\overline{(\Delta\alpha^2)}\frac{R^2}{t}. \tag{14}$$

In Eq. (14) $b$ is a constant number whereas $R$ and $t$ are characteristic length and the time, respectively. Here $R$ is the maximum radius of the vessels

From Eq.(12) we have

$$\frac{l^2}{t} = b\frac{t}{\Delta\tau}\alpha^2\frac{R^2}{t}. \tag{15}$$

Through Eq. (15) we obtain the diffusion length, or what is the same in our case, the correlation length,

$$L = \alpha R \left(b\frac{t}{\Delta\tau}\right)^{\frac{1}{2}}. \tag{16}$$

(ii)To compute the contribution coming from a cluster of correlated vessels we perform a simple sum of currents coming from every one of four (in a cluster of first order) uncorrelated vessels. The resultant current is them divided by four (the order of the cluster) such we have just the mean current for each vessel in the cluster. This mean current corresponds to a cluster originated from one of the original nucleation points, such the total number $N$, used to compute the total current in the initial time, still remains. The total current is always given by the formula:

$$I_t(t) = N\left[\left(\frac{L_d^{(n)} - L}{L_d^{(n)}}\right)I^{(n-1)} + \left(\frac{L}{L_d^{(n)}}\right)I^{(n)}\right], \quad n = 1,2,3.., \tag{17}$$

where $I^{(n-1)}$ and $I^{(n)}$ are the mean currents in a cylinder that composes a cluster of order $n-1$ and $n$, respectively, and the terms in parenthesis are the weight factors that answer for their relative contributions, which corresponds to item (iii) listed before. In Eq. (17) $L_d^{(n)}$ is the length corresponding to the correlation of vessels in an $n$ order cluster (just a geometrical measure), whereas $L$ is the time dependent correlation length given by Eq. (16), and $I^{(0)}$ is the current flowing in a single vessel. In Fig. 5 we show a straight line corresponding to $L_d^{(1)} = 1.154R$.

Using the corrections introduced in Eq. (10) to compute Eq. (6), and also Eq. (17) to compute the total current, we obtain the (b) profile shown in Fig. 6. It is just this current transient which must be compared to those shown by Sapoletova et al. [6] (Fig. 4 in their article).

The attenuation of the oscillations around the stationary value of the current depends on the value assigned to the constant $b$ appearing in eq. (16). In Fig. 6 we used $b = 0.29$ only to get a better agreement with the results presented by Sapoletova [6]. Given the equality of the statistics that define the parameter $\alpha$, and also the diffusivity, it would be natural to find $b = 4$, but we may observe that the fluctuating forces (that define diffusivity and come from the unbalanced molecular forces) are homogeneously distributed in the space, showing the same behavior everywhere, whereas the fluctuations on $\alpha$ are not homogeneous in space, although our model conceives them as regularly distributed in the space.

In the inset of Fig. 6 we show a comparison between the currents flowing in one uncorrelated vessel (black line) and that flowing through a single vessel but now correlated to three others, in a cluster of order four (the gray line).

## 4. CONCLUSIONS

In this paper we demonstrated that an ordered porous system can be represented by a set of cylindrical vessels with permeable walls. A set of boundary and initial conditions for diffusion and reaction in a cylindrical vessel results in an analytical expression for the current transient measured at the bottom of the cylinder that is completely defined by a group of parameters that control the intensity of the ionic flux through the walls and the rate of ion consumption at the bottom of the cylinder. The introduction of a periodic corrugation in the cylindrical was essential to reproduce the current minima observed in experiments as well as the assumption of a dissipative process that follows the random fluctuations in the path's line. The model is able to decouple the geometric features of the system from the reaction-diffusion kinetics and succeed to show that the current yield decrease with the inverse of the porous layer thickness.

## ACKOWLEDGEMENT

We wish to acknowledge our obligation to Dr. Wagner Figueiredo for improvements of phrases and construction, and, above all for his criticism.

**FIGURE CAPTIONS**

**Figure 1**: (a) Schematic representation of the flux through a porous system and the stream lines in a give instant of time. Some of the stream lines enter the duct bounded by the thick lines in (b) whereas another one leaves the conduct at different places. In the next time the place and direction of the stream lines that enter or leave the duct change.

**Figure 2**: (a) Top view of a colloidal crystal formed by four layers of spheres self-ordered on a flat substrate in a fcc structure. The star at the center indicates one pore through which an impinging

ion may enter the porous structure. The write arrows depict one possible diffusion path towards the flat substrate. (b) Schematic cross section of the colloidal crystal. Black arrows indicate the diffusion path that was singled out in (a). (c) That particular diffusion path is now modeled as a staircased cylinder. The whole porous structure can be seen as a periodic replication of aligned twisted vessels. At each inflection point the vessels may exchange particles with neighboring ones. (d) In the last simplification step, the twisted vessels transform into straight cylinders with corrugated walls. The function $g(z)$ is sketched on the right, indicating the points of maximum and minimum flux among vessels.

**Figure 3**: Current transients for one single vessel when the convective velocity is changed from positive to negative values as indicated in the figure. To perform the calculations we used $D = 1x10^{-6} cm^2/s$, $c_b = 26\,mM$, $h = 3x10^{-3}cm$, $v = 0.1s^{-1}$, $\beta = 150.$, $k = 9.89s^{-1}$ and $R = 6x10^{-5}cm$.

**Figure 4**: Theoretical current transients for a four layer medium without convective velocity. Curve (a) is identical to that showed in figure 3, when $v_c = 0$, and there is no corrections in the magnitude of $h$. Curve (b) is a similar profile, obtained when corrections are introduced in the value of $h$. At the inset we show two current profiles, with $h$ corrections; the black one for a four layer system, and the gray corresponding to a template constituted by just one monolayer of spheres. In spite of the little difference among the systems we observe a shift of the black curve toward lower current yield. All parameters are defined by the value depicted in figure 3 except the $h$ corrections.

**Figure 5**: Schematic representation of a system formed by one layer of spheres ordered on a flat electrode. Top view indicates in (a) twelve nucleation points with a growth hemisphere of deposits under development in its early stages. Figure (b) represents the same system in a posterior time if compared to (a). The twelve nucleation point show hemispheres in advanced stages, two of them are showing correlated flux with neighbouring vessels. Figure (c) shows another instant of time, where the group of correlated vessels increases to four. Figure (d) shows another situation where eight groups of clusters of order one is present. In figure (a) we indicate with an arrow the magnitude of $L_d^{(1)}$ that enters into equation (17).

**Figure 6**: Theoretical current transients. The gray curve is identical to those shown in figure 3, and also in figure 4 (curve (a) in both), which means it does not include the correction in $h$ and also the effect of correlation among vessels, once it is a profile for a single vessel. The black curve is the total current divided by the number of nucleation points in order to be compared with that for one single vessel (the gray one). The black curve was obtained using Eq. (10) and (17)> The parameters are the same as indicated in figure 3, except $h$ and, $v_c = 0$ here. In the inset, we show the current profile for a single not correlated vessel (gray curve) and the current profile for a single vessel correlated to another three of them.

**APPENDIX**

In this appendix we will show how to derive the expression for the current given by Eq. (3) from the following differential equation,

$$\frac{\partial}{\partial t} C(r,z,t) = D\left[\frac{1}{r}\frac{\partial}{\partial r}\left(r\frac{\partial}{\partial r}\right) + \frac{\partial^2}{\partial z^2}\right] C(r,z,t) + v_c \frac{\partial}{\partial z} C(r,z,t) \,, \tag{A1}$$

and the boundary and initial conditions,

$$C(r,z,0) = c_b, \tag{A2a}$$

$$C(r,0,t) = (c_b - c_s)e^{-kt} + c_s, \tag{A2b}$$

$$C(r,h,t) = c_b, \tag{A2c}$$

$$\left(\frac{\partial c}{\partial r}\right)_{r=0} = 0, \tag{A2d}$$

$$\left(\frac{\partial c}{\partial r}\right)_{r=R} = -\alpha c_b (1 - e^{-\nu t}) g(z). \tag{A2e}$$

In order to simplify Eq.(A1) we perform a variable transformation given by:

$$C(r,z,t) = U(r,z,t) exp\left(\frac{v_c}{2D}z - \frac{v_c^2}{4D}t\right), \tag{A3}$$

which reads like,

$$C(r,z,t) = U(r,z,t) e^{\beta z - \Omega t}, \tag{A4}$$

after definitions: $\frac{v_c}{2D} = \beta$, and $\frac{v_c^2}{4D} = \Omega$. Expression (A3) transforms Eq. (A1) into another equation,

$$\frac{\partial}{\partial t} U(r,z,t) = D\left[\frac{1}{r}\frac{\partial}{\partial r}\left(r\frac{\partial}{\partial r}U\right) + \frac{\partial^2}{\partial z^2}U\right]. \tag{A5}$$

In a similar way, the boundary and initial conditions transforms to:

$$U(r,z,0) = c_b e^{-\beta z}, \tag{A6a}$$

$$U(r,0,t) = (c_b - c_s)e^{-kt+\Omega t} + c_s e^{\Omega t}, \tag{A6b}$$

$$U(r,h,t) = c_b e^{-\beta h + \Omega t}, \tag{A6c}$$

$$\left(\frac{\partial U}{\partial r}\right)_{r=0} = 0, \tag{A6d}$$

$$\left(\frac{\partial U}{\partial r}\right)_{r=R} = -\alpha c_b (1 - e^{-\nu t}) g(z) e^{-\beta z + \Omega t}. \tag{A6e}$$

Because we search for a convenient form to compare the theoretical data and the experimental ones we compute the charge current that crosses the reactive surface in its normal direction z. This current is the measured quantity in experimental realizations and, in our model it is implemented through the equation:

$$J(r,z,t) = -D\bar{z}F\frac{\partial}{\partial z}C(r,z,t), \tag{A7}$$

the component of the current density along the z direction. $\bar{z}$ is the charge number, connected to the number of electrons withdrawn from the electrode to reduce the ionic species on its surface and $F$ is the Faraday constant. In this sense the integral of Eq. (A7), performed on the electrode surface, give us the current charge that crosses the reactive surface, that is:

$$I(z,t) = \int_0^R 2\pi r J(r,z,t) dr = -D\bar{z}F\frac{\partial}{\partial z}\int_0^R 2\pi r\, C(r,z,t) dr, \tag{A8}$$

and, using Eq.(A4),

$$I(z,t) = -D\bar{z}F\frac{\partial}{\partial z,}\Big(e^{\beta z - \Omega t}\int_0^R 2\pi r\, U(r,z,t)dr\Big),$$ (A9)

which could be write as,

$$I(z,t) = -D\bar{z}F\frac{\partial}{\partial z}\Big[e^{\beta z - \Omega t}u(z,t)\Big],$$ (A10)

Were we defined

$$u(z,t) = \int_0^K 2\pi r\, U(r,z,t)dr.$$ (A11)

Now, through these definitions, we could multiply Eq. (A5) by $2\pi r$, before integrating it in the $r$ variable, in order to obtain a new and simpler equation,

$$\frac{\partial}{\partial t}u(z,t) = -2\pi R D\alpha g(z)c_b H(t)e^{-\beta z + \Omega t} + D\frac{\partial^2}{\partial z^2}u(z,t),$$ (A12)

where we have defined,

$$H(t) = (1 - e^{-vt}).$$ (A13)

Also, the boundary and initial conditions change by use of Eq. (A11), such:

$$u(z,0) = \pi R^2 c_b e^{-\beta t},$$ (A14a)

$$u(0,t) = \pi R^2(c_b - c_s)e^{-kt + \Omega t} + \pi R^2 c_s e^{\Omega t},$$ (A14b)

$$u(h,t) = \pi R^2 c_b e^{-\beta h + \Omega t}.$$ (A16c)

Our original problem now is reduced to a simpler equation (A12) but still has non homogeneous boundary and initial conditions. Before going further, we introduce a more compact notation, so we write:

$$u_t - Du_{zz} = q(z,t),$$ (A17)

$$u(0,t) = A(t),$$ (A18a)

$$u(h,t) = B^{(1)}(t),$$ (A18b)

$$u(z,0) = B^{(2)}(z).$$ (A18c)

An additional transformation makes homogeneous the boundary conditions, allowing us to use what is called the method of variation of parameters [15]. We choose a continuous and differentiable function $K(z,t)$ defined by,

$$K(z,t) = \frac{z}{h}B^{(1)}(t) - \frac{z-h}{h}A(t),$$ (A19)

such $u(z,t)$, can be written as

$$u(z,t) = v(z,t) + K(z,t),$$ (A20)

and, $v(z,t)$ is also a continuous and differentiable function of its variables.

Through the application of Eq.(A20) into differential equation (A17) we get,

$$v_t(z,t) - Dv_{zz}(z,t) = Q(z,t),$$ (A21)

where,

$$Q(z,t) = q(z,t) - \frac{z}{h}B_t^{(1)} + \frac{z-h}{h}A_t(t),$$ (A22)

with the boundary and initial conditions,

$$v(0,t) = u(0,t) - K(0,t) = 0,$$ (A23a)

$$v(h,t) = u(h,t) - K(h,t) = 0,$$ (A23b)

$$v(z,0) = u(z,0) - K(h,0) = B^{(2)}(z,0) - \frac{z}{h}B^{(1)}(0) + \frac{z-h}{h}A(0) = f(z).$$ (A23c)

Now we have a partial differential equation with homogeneous boundary conditions. The method of variation of parameters [15] assumes solutions in the form,

$$v(z,t) = \sum_{n=1}^{\infty} T_n(t)\varphi_n(z),$$ (A24)

where, $\varphi_n(z)$ are the eigenfunctions of the related homogeneous problem, i.e.:

$$\varphi_{n(z)} = sin(\omega_n z),$$ (A25)

and $\omega_n = \frac{n\pi}{h}$ is the correspondent eigenvalue. Because the $\varphi_n(z)$ eigenfunctions are a set of orthogonal functions, multiplying Eq. (A24) by $\varphi_m(z)$, followed by an integration on the $z$ variable, gives

$$T_n(t) = \frac{2}{h}\int_0^h v(z,t)\varphi_n(z)dz.$$ (A26)

Taking the time derivative of Eq. (A26), and using the expression (A21) we get,

$$T_n'(t) = \frac{2}{h}\int_0^h Dv_{zz}\varphi_n(z)dz + Q_n(t),$$ (A27)

where,

$$Q_n(t) = \frac{2}{h}\int_0^h Q(z,t)\varphi_n(z)dz,$$ (A28)

and $T_n'(t)$ means the first derivative with respect to time.

Using the Green formula,

$$\int_0^h v_{zz}\varphi_n(z)dz = [v_z\varphi_n - v\varphi_n']_0^h + \int_0^h v\varphi_n''dz, \tag{A29}$$

and remembering that $v(z,t)$, as well as $\varphi_n(z)$ fulfill homogeneous boundary conditions given by

$$\varphi_n'' = -\omega_n^2\varphi_n, \tag{A30}$$

we get

$$T_n'(t) = -D\omega_n^2 T_n(t) + Q_n(t), \tag{A31}$$

where  use was made of Eq. (A26). The integration of Eq. (A31) turns to be an easy task after a multiplication by $e^{D\omega_n^2 t}$ , so,

$$T_n(0) = \frac{2}{h}\int_0^h f(z)\varphi_n(z)dz = c_n, \tag{A32}$$

$$T_n(t) = e^{-D\omega_n^2 t}c_n + e^{-D\omega_n^2 t}\int_0^t Q_n(s)e^{D\omega_n^2 s}ds, \tag{A33}$$

and as a consequence,

$$v(z,t) = \sum_{n=1}^{\infty} sin(\omega_n z)\left[c_n e^{-D\omega_n^2 t} + e^{-D\omega_n^2 t}\int_0^t Q_n(s)e^{D\omega_n^2 s}ds\right]. \tag{A34}$$

To recover $u(z,t)$ we must invoke Eq. (A20), such

$$Q_n(s) = \frac{2}{h}\int_0^h\left[q(z,s) - \frac{z}{h}\frac{\partial}{\partial t}B^{(1)} + \frac{z-h}{h}\frac{\partial}{\partial t}A\right]sin(\omega_n z)\,dz, \tag{A35}$$

and

$$c_n = \frac{2}{h}\int_0^h\left[B^{(2)}(z,0) - \frac{z}{h}B^{(1)}(0) + \frac{z-h}{h}A(0)\right]sin(\omega_n z)\,dz. \tag{A36}$$

Finally, using the definitions that appear in (A18), we arrive at,

$$u(z,t) = \frac{z}{h}\pi R^2 c_b e^{-\beta h+\Omega t} - \frac{z-h}{h}\pi R^2\left[\pi R^2(c_b-c_s)e^{-kt+\Omega t} + \pi R^2 c_s e^{\Omega t}\right] - \frac{2}{h}\sum_{n=1}^{\infty}\frac{\pi R^2 c_b\beta^2}{\beta^2-\omega_n^2}\left(1 - cos(n\pi)e^{-D\omega_n^2 t}\right)sin(\omega_n z) -$$
$$\frac{2}{h}\sum_{n=1}^{\infty} 2\pi R D\alpha c_b g_n\left[\frac{e^{\Omega t}-e^{-D\omega_n^2 t}}{\Omega+\omega_n^2 D} - \frac{e^{(\Omega-v)t}-e^{-D\omega_n^2 t}}{\Omega+\omega_n^2 D-v}\right]sin(\omega_n z) +$$
$$\frac{2}{h}\sum_{n=1}^{\infty}\frac{\pi R^2}{\omega_n}c_b\Omega cos(n\pi)e^{-\beta h}\frac{\left(e^{\Omega t}-e^{-D\omega_n^2 t}\right)}{\Omega+D\omega_n^2}sin(\omega_n z) - \frac{2}{h}\sum_{n=1}^{\infty}\frac{\pi R^2}{\omega_n}(c_b-c_s)\left(\frac{\Omega-k}{\Omega-k+D\omega_n^2}\right)\left(e^{(-k+\Omega)t} - e^{-D\omega_n^2 t}\right)sin(\omega_n z) + \sum_{n=1}^{\infty}\frac{\pi R^2}{\omega_n}c_s\Omega\left(\frac{1-cos(n\pi)}{\Omega+D\omega_n^2}\right)\left(e^{\Omega t}-e^{-D\omega_n^2 t}\right)sin(\omega_n z)$$
$$. \tag{A37}$$

In Eq. (A37), $g_n = \int_0^h g(z)\,sin(\omega_n z)\,dz$ .

Using Eq. (A9), we obtain the total current that flows through the reactive base of the cylindrical cavity, when z=0:

$$I(0,t) = -\bar{z}FD\pi R^2 \left\{ (c_b - c_s)e^{-kt}\left(\beta - \frac{1}{h}\right) + c_s\left(\beta + \frac{1}{h}\right) + \frac{c_b}{h}e^{-\beta h} + \frac{2}{h}\sum_{n=1}^{\infty}\frac{c_b\beta^2\omega_n}{\beta^2+\omega_n^2}\left(\cos(n\pi)e^{-\beta h} - 1\right)e^{-\mu_n t} + \frac{2}{h}\sum_{n=1}^{\infty}\frac{c_b\Omega\cos(n\pi)e^{-\beta h}}{\mu_n}(1 - e^{-\mu_n t}) + c_s\Omega\sum_{n=1}^{\infty}\left(1 - \cos(n\pi)\frac{(1-e^{-\mu_n t})}{\mu_n}\right) + \frac{2}{h}(c_b - c_s)(\Omega - k)\sum_{n=1}^{\infty}\frac{(e^{-\mu_n t}-e^{-kt})}{\mu_n - k} + \frac{2}{h}\frac{D\alpha c_b}{R}\sum_{n=1}^{\infty}2g_n\omega_n\left[\frac{(e^{-\nu t}-e^{-\mu_n t})}{\mu_n - \nu} - \frac{(1-e^{-\mu_n t})}{\mu_n}\right]\right\}$$

,                    (A38)

where

$$\beta = \frac{v_c}{2D},$$

$$\Omega = \beta^2 D,$$

and

$$\mu_n = \frac{Dn^2\pi^2}{h^2} + \frac{v_c^2}{4D}.$$

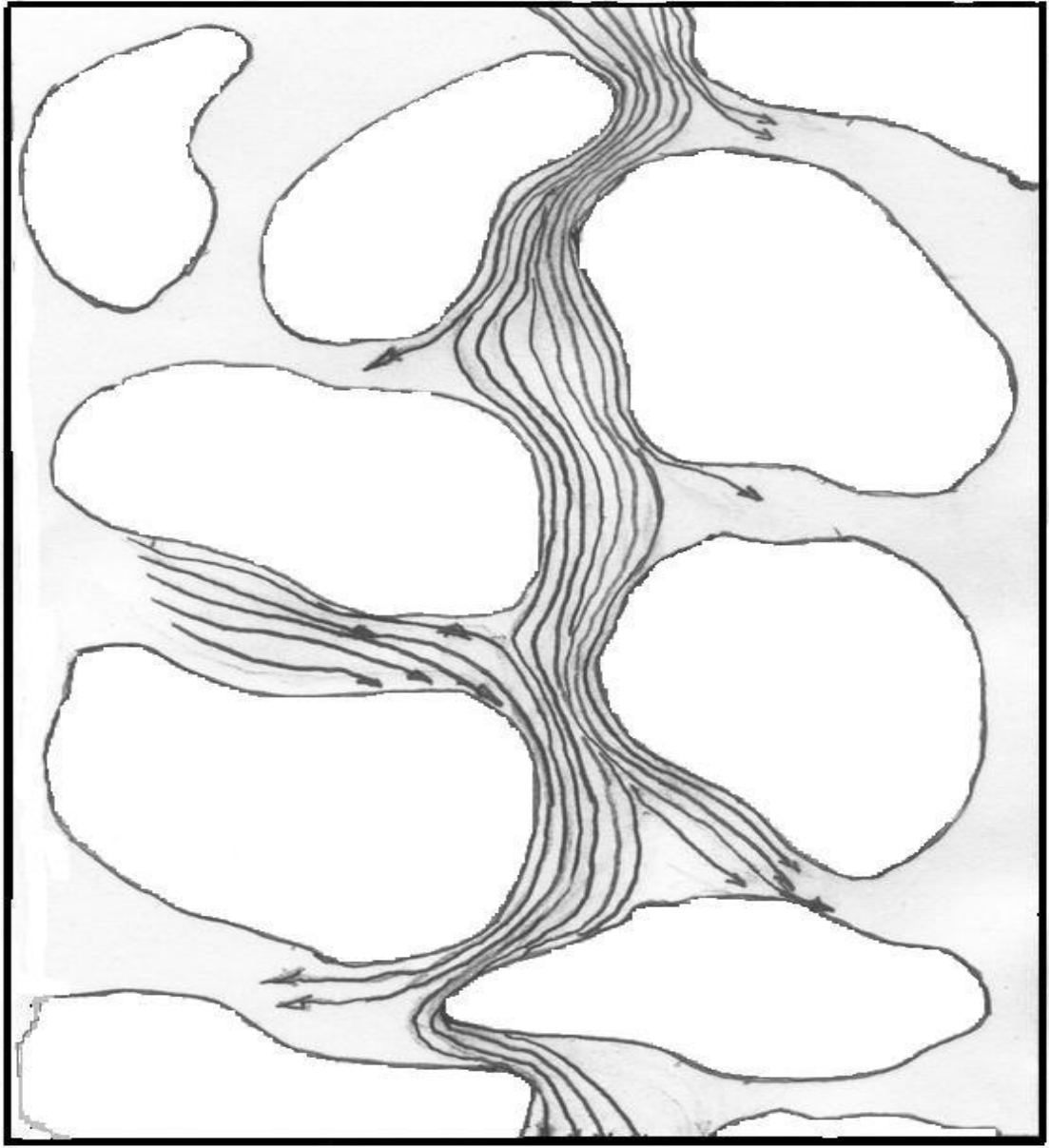

*Figure 1*

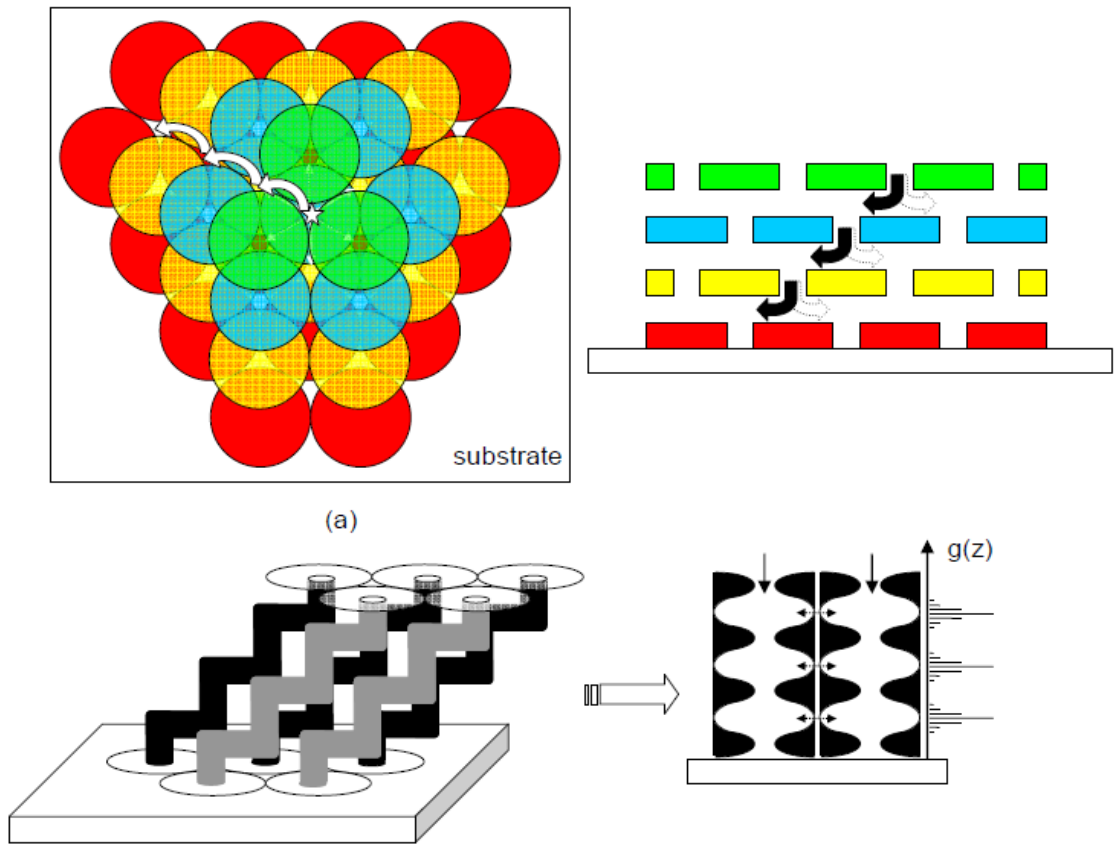

(a)

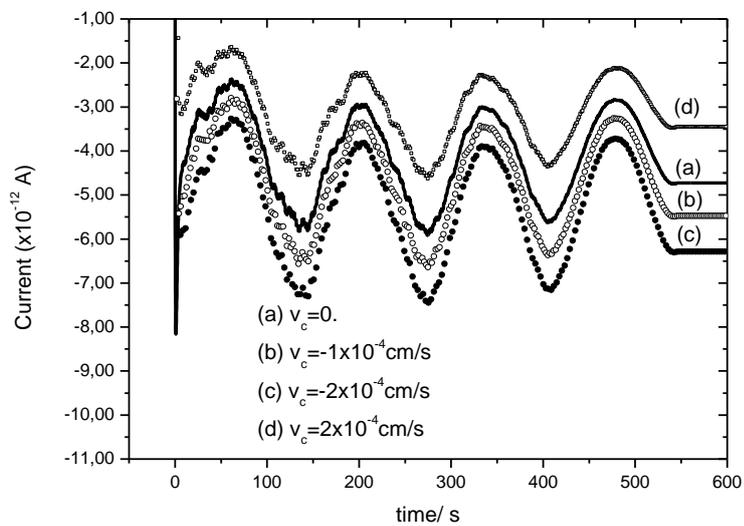

*Figure 2*

Current (x10⁻¹² A) vs time/ s

(a) $v_c$=0.
(b) $v_c$=-1x10⁻⁴cm/s
(c) $v_c$=-2x10⁻⁴cm/s
(d) $v_c$=2x10⁻⁴cm/s

*Figure 3*

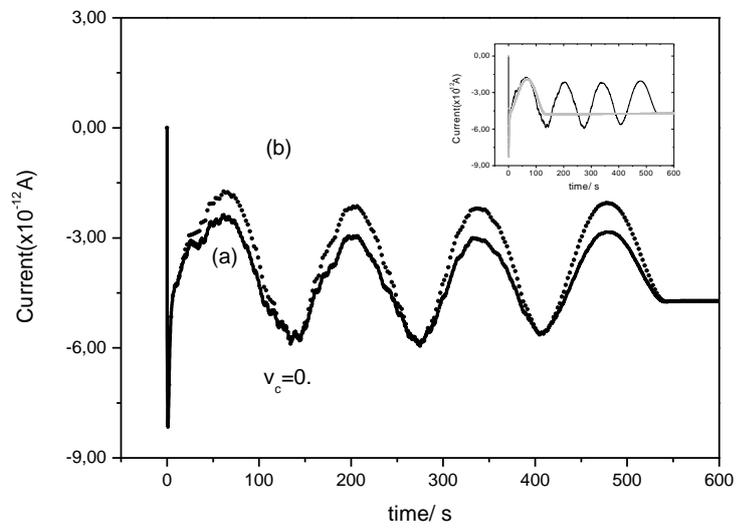

*Figure 4*

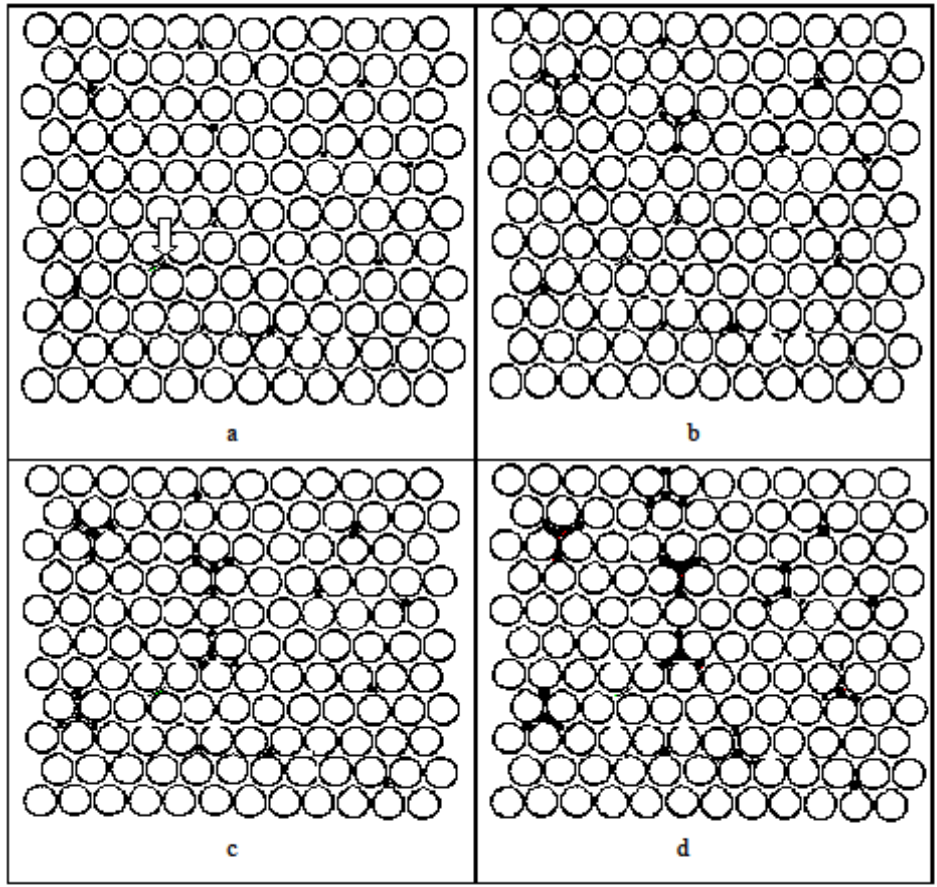

Figure: 5

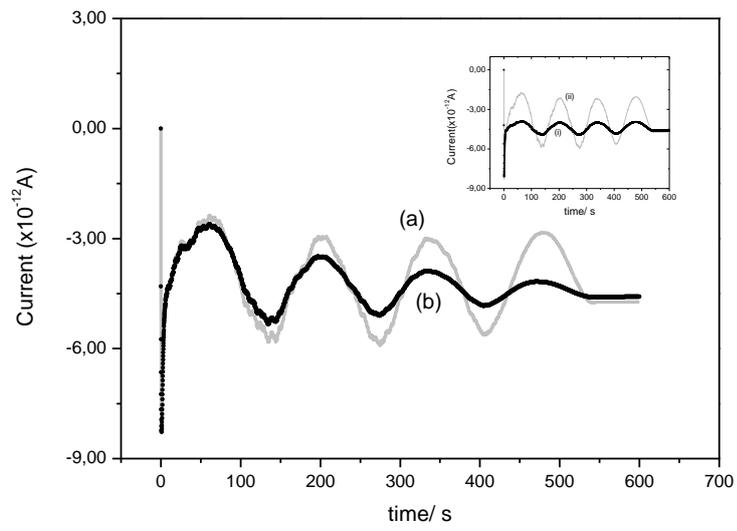

*Figure 6*